\documentclass{aa}
\usepackage{color}
\usepackage{amsmath}
\usepackage{float}
\usepackage{natbib}
\usepackage{graphicx}
\usepackage{graphicx}
\usepackage{ulem}
\usepackage[varg]{txfonts}
\usepackage{lscape}
\usepackage{xspace}

\definecolor{magenta}{rgb}{1.0, 0.0, 1.0}
\definecolor{mygray}{gray}{0.6}

\newcommand{\eq}[1]{Eq.\,(\ref{eq:#1})}

\def\me{$\,{\rm M}_{\oplus}$\xspace}

\def\h2o{H$_2$O}
\def\sio2{SiO$_2$}
\def\kgm3{kg\,m$^{-3}$}
\def\Jm3{J\,m$^{-3}$}
\def\Jkg3{J\,kg$^{-3}$}
\def\Jkg1{J\,kg$^{-1}$}
\def\cd{$C_\textrm{d}\,$}

\begin{document} 

   \title{How cores grow by pebble accretion}
   \subtitle{I. Direct core growth}

   \author{M. G. Brouwers
          \and
          A. Vazan 
          \and
          C. W. Ormel 
          }

   \institute{Anton Pannekoek Institute, University of Amsterdam, Science Park 904, PO box 94249, Amsterdam, The Netherlands \\
              \email{mgbrouwers@gmail.com; A.Vazan@uva.nl; c.w.ormel@uva.nl}
             }

  \abstract
{Planet formation by pebble accretion is an alternative to planetesimal-driven core accretion. In this scenario, planets grow by the accretion of cm- to m-sized pebbles instead of km-sized planetesimals. One of the main differences with planetesimal-driven core accretion is the increased thermal ablation experienced by pebbles. This can provide early enrichment to the planet's envelope, which influences its subsequent evolution and changes the process of core growth.}
{We aim to predict core masses and envelope compositions of planets that form by pebble accretion and compare mass deposition of pebbles to planetesimals. Specifically, we calculate the core mass where pebbles completely evaporate and are absorbed before reaching the core, which signifies the end of direct core growth}
  {We model the early growth of a protoplanet by calculating the structure of its envelope, taking into account the fate of impacting pebbles or planetesimals. The region where high-Z material can exist in vapor form is determined by the temperature-dependent vapor pressure. We include enrichment effects by locally modifying the mean molecular weight of the envelope.}
   { In the pebble case, three phases of core growth can be identified. In the first phase ($M_\textrm{core}$ < 0.23--0.39 \me), pebbles impact the core without significant ablation. During the second phase ($M_\textrm{core}$ < 0.5 \me), ablation becomes increasingly severe. A layer of high-Z vapor starts to form around the core that absorbs a small fraction of the ablated mass. The rest of the material either rains out to the core or instead mixes outwards, slowing core growth. In the third phase ($M_\textrm{core}$ > 0.5 \me), the high-Z inner region expands outwards, absorbing an increasing fraction of the ablated material as vapor. Rainout ends before the core mass reaches 0.6 \me, terminating direct core growth. In the case of icy \h2o pebbles, this happens before 0.1 \me.}
   {Our results indicate that pebble accretion can directly form rocky cores up to only 0.6 \me, and is unable to form similarly sized icy cores. Subsequent core growth can proceed indirectly when the planet cools, provided it is able to retain its high-Z material.}


   \maketitle
%

\section{Introduction}
The growth of planetary cores is a conceptually simple process in the planetesimal-driven core accretion scenario. An increasingly massive planetary embryo is impacted by km-sized impactors that add to its mass \citep{Pollack1996, Hubickyj2005}. Soon, the small protoplanet starts to gravitationally bind an envelope of gas. This envelope is initially very poor in high-Z material due to the limited interaction between the envelope and any impacting planetesimals \citep{Podolak1988}. Core growth eventually comes to a halt when the planet has no more solid material in its feeding zone left to accrete. Unfortunately, the planetesimal-driven core accretion scenario faces a time-constraint problem at distances beyond several AU. Simulations indicate that planet formation typically requires more time than the expected lifetimes of the disks from which their matter originates \citep{Kobayashi2010, Levison2010, Bitsch2015}. The underlying problem is the strong dynamical excitation a growing planet induces to the planetesimal disk, i.e., a negative feedback effect. Although planetesimal-driven planet formation scenarios may still be viable with optimal choices regarding planetesimal sizes and strengths \citep[e.g.,][]{KobayashiEtal2016}, an alternative approach is to investigate formation models that are not affected by this problem.

One such scenario is pebble accretion \citep{Ormel2010, Lambrechts2012}. Instead of km-sized planetesimals, the protoplanet accretes smaller objects (cm- to m-sized) called pebbles. There are two main differences that originate from this size difference. Firstly, pebble accretion is expected to accrete mass approximately an order of magnitude more rapidly than planetesimals \citep{Lambrechts2014}, and operates particularly well when the pebbles are settled into the mid-plane regions \citep{Ormel2017}. The reason for this is the more efficient capturing of pebbles due to their increased susceptibility to gas drag. Secondly, pebbles feature a stronger interaction with a planet's atmosphere. Their higher area-to-mass ratio makes it easier for them to be slowed down by aerodynamic gas drag and reduces their ablation timescales \citep[e.g.,][]{ Love1991, Mcauliffe2006}. As a result, these pebbles can deposit high-Z vapor into even a small growing planet. This in turn can have a large effect on the planet's further growth and evolution \citep{Venturini2016}. The presence of high-Z material locally increases the mean molecular weight, which drives up densities and temperatures, increasing the envelope mass.

So far, research on pebble accretion has mainly focused on the first difference by calculating accretion rates though disk evolution and drag considerations \citep[e.g.,][]{Morbidelli2012, Chambers2014, Ida2016b}. The evaporation effects have only recently been considered by \citet{Alibert2017}, based on impact simulations by \citet{Benz2006}. In his simulations, direct core impacts terminate either when the planet becomes too hot ($T_\textrm{core} > 1600 $ K) or the envelope becomes too massive ($M_\textrm{env} > 10^{-4}$ \me) for pebbles to reach the core. He found these conditions to be met before the core grew to 1 \me. However, he neglected any enrichment effects and did not compute the planet's evolution beyond the point of full ablation. In reality, the core can continue to grow if ablated material oversaturates the gaseous atmosphere, and the leftover material rains out to the core \citep{Iaroslavitz2007}.

In this work, we investigate how cores grow by pebble accretion. We expand previous calculations of core growth by including enrichment effects from when they first occur and by considering further growth through the rainout of ablated material. Our code simulates planet formation along with pebble impacts in a way that quantitatively incorporates how the pebbles interact with the growing atmosphere. We calculate the effects of ablation and gas drag on impacting pebbles and relate these effects to the planet's subsequent evolution by modifying the luminosity profile and mass deposition curve. Our results show that impacting pebbles will be fully ablated before the planet even reaches 0.5 \me. Subsequent rainout of ablated material in a super-saturated atmosphere can add $\sim$0.1 \me to the core mass, but this direct core growth terminates well before the core mass has reached $1\,$\me.

This paper is organized as follows. Sections \ref{Impact model} and \ref{Planetary structure} introduce our impact and planet simulation models. Section \ref{High-Z enrichment and rainout} elaborates on our treatment of high-Z enrichment and on the rainout of ablated material to the core. Our pebble simulation results for \sio2 and \h2o are presented in Sects. \ref{Standard model} -- \ref{Icy pebbles}. We compare planetesimal mass deposition in Sect. \ref{Mass deposition} and examine the core-surface conditions in Sect. \ref{Core surface conditions}. Section \ref{Discussion and conclusions} covers the discussion and conclusions. 

\section{Model description}
The goal of this work is to provide an estimate of the core sizes that planets can form directly by pebble accretion. To clarify our approach, we distinguish three distinct mechanisms by which planetary core growth can occur. Direct impact of solids reaching the core, the rainout of ablated material if the envelope is super-saturated, and a phase change of gas in the inner envelope under sufficiently high pressure. We consider only the first two mechanisms in this work, and show that the third is not yet relevant for our early model. This is what we mean by direct core growth. We limit ourselves to the solids accretion phase when the planet's accretion is still dominated by solids instead of disk gas and Kelvin--Helmholtz contraction is not yet significant.

In our formation scenario, a growing protoplanet is constantly being impacted by small, cm- to m-sized objects. These impactors partially or fully evaporate during entry and contribute high-Z material (in our case \sio2, quartz) to the core and envelope. To simulate the planet's growth, we used a new evolution code consisting of two components. The first is a time-dependent calculation of the planet's structure and gas accretion rate. We modeled its evolution, as is typically done, by a series of quasi-hydrostatic models with a separate core and envelope \citep[e.g.,][]{Mordasini2012,Piso2014}. The second component of our code is a calculation of the interaction between impacting pebbles and the planet's atmosphere. The accretion of solids is modeled by a single-body simulation code that uses the layers generated by the structure code. Impactors hit the planet head-on and lose mass by thermal ablation and friction. The resultant core accretion and mass deposition rates affect the planet's further evolution. In principle, the two codes should be iterated to make sure they work self-consistently. In tests, one iteration was found to be sufficient to achieve this. Therefore, we ran them twice during every time step.

\subsection{Impact model}\label{Impact model}

\subsection*{Gas drag}\label{Gas drag}
\begin{figure}[h!] 
\centering
\includegraphics[width=\hsize]{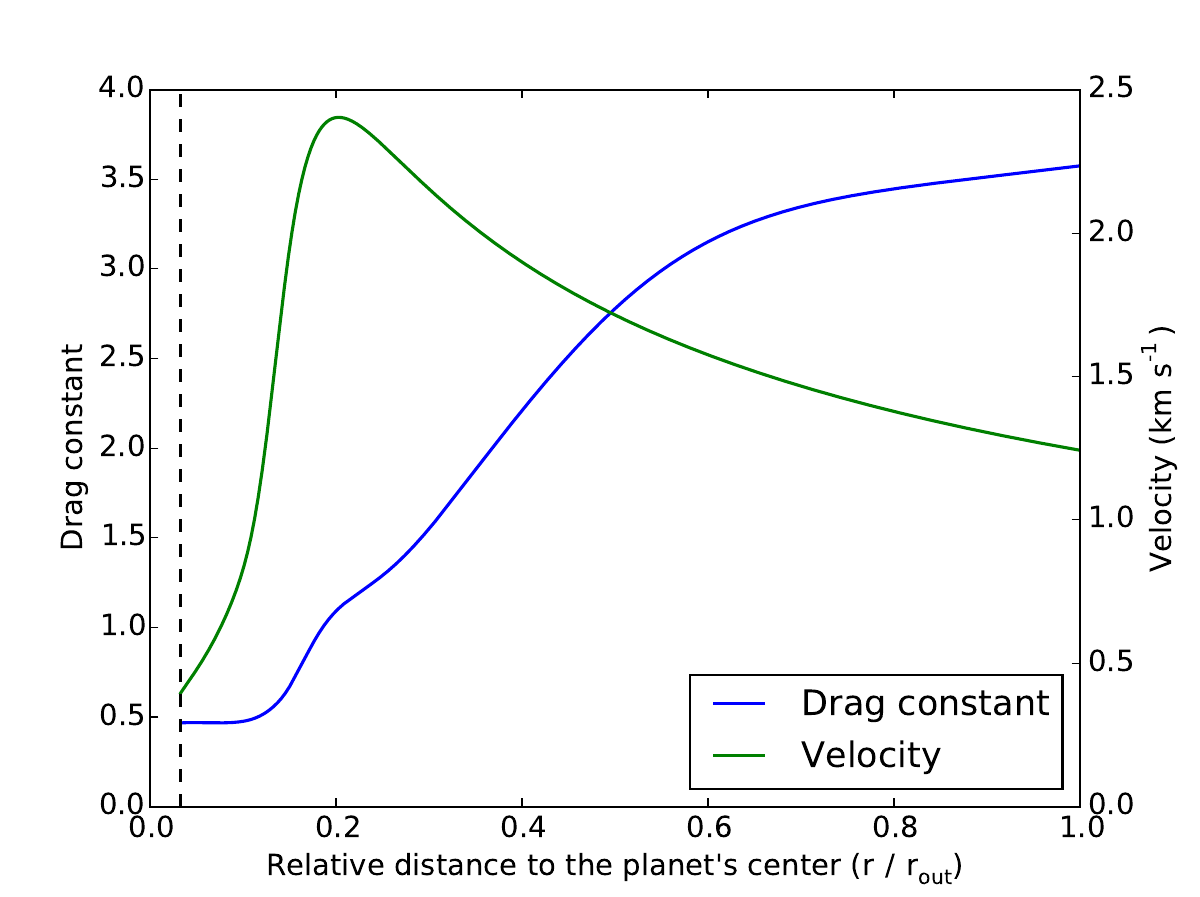} 
\caption{ Drag constant evolution of a 0.1 m \sio2 pebble impacting a 0.3 \me planet at escape velocity. The blue line represents the drag constant, the green line indicates the impactor's velocity relative to the surrounding gas, and the black dashed curve represents the core radius. \label{drag_evolution}} 
\end{figure}

The pebbles are modeled as perfect spheres of uniform size and a constant density of 2.65 $\times \, 10^3$ \kgm3. We further assume that they impact the planet head-on. Pebbles that lack spherical symmetry or enter the planet's atmosphere at an angle will experience additional ablation due to their increased travel time. In this sense, we study the limit in which pebbles can most easily reach the core. Their initial velocities are set to the lower of the terminal and escape velocities. The first is a drag-induced limitation, calculated by the velocity at which the local drag force is exactly equal in magnitude to the gravitational force. The second is the limitation of the total gravitational energy that an impactor can acquire as it moves towards the planet. As impactors approach the core, gravity accelerates them and gas drag slows them down. The drag force is given by
\begin{equation}
F_\textrm{drag} = 0.5 C_\textrm{d} A_\textrm{i} \rho_\textrm{g} v_\textrm{i}^2
,\end{equation}
where $\rho_\textrm{g}$ is the gas density, $A_\textrm{i}$ the impactor's frontal area, $v_\textrm{i}$ its relative velocity to the gas, and \cd the drag constant. Most of the physics is contained in this drag constant. For our purposes, we want to be able to describe drag over a wide range of conditions for impactors varying in size from 1 cm to several kilometers. There has been extensive research on empirical expressions for different drag regimes. We adopt an extended version of the general expression by \citet{Melosh2008} based on a review of this literature. Theirs is a continuous expression of \cd that is applicable to a wide range of conditions. It depends on the local mach number $\textrm{Ma} = v_\textrm{i} / c_\textrm{s, g}$, where $c_\textrm{s, g}$ is the local sound velocity; on the Reynolds number $\textrm{Re} = v_\textrm{i} R_\textrm{i} / \eta_\textrm{g}$, where $\eta_\textrm{g}$ is the dynamical viscosity of the gas and $R_\textrm{i}$ is the impactor radius; and on Knudsen's number (Kn), which is defined as the ratio between the Mach and Reynolds numbers, $\textrm{Kn} = \frac{\textrm{Ma}}{\textrm{Re}}$. The full expression for $C_\textrm{d}$ reads as
\begin{equation}
 C_\textrm{d} = 2 + \left(C_1 - 2\right) \, \textrm{exp}[-3.07 \sqrt{\gamma_\textrm{g}} \textrm{Kn} C_2] + H \frac{1}{\sqrt{\gamma_\textrm{g}} \textrm{Ma}} e^{-1 / 2 \textrm{Kn}}
,\end{equation}
where $C_\textrm{1}$, $C_\textrm{2}$, and $H$ are the auxiliary functions
\begin{equation}
C_1 =\frac{24}{\textrm{Re}} \left(1 + 0.15 \textrm{Re}^{0.678}\right) + \frac{0.407 \textrm{Re}}{\textrm{Re} + 8710}
\end{equation}
\begin{equation}
\log C_2 = \frac{2.5 \left(\textrm{Re} / 312\right)^{0.6688}}{1 + \left(\textrm{Re} / 312\right)^{0.6688}}
\end{equation}
\begin{equation}
H = \frac{4.6}{1 + \textrm{Ma}} + 1.7 \sqrt{\frac{T_\textrm{i}}{T_\textrm{g}}}
\end{equation}
and $T_\textrm{s}$ and $T_\textrm{g}$ are the impactor's surface temperature and the local gas temperature, respectively; $\gamma_\textrm{g}$ is the adiabatic index of the gas, explained in detail in Sect. \ref{Planetary structure}; and $C_\textrm{1}$ is an extended version of the expression given by \citet{Melosh2008} as suggested by \citet{Podolak2015}, based on an empirical review from \citet{Brown2003}. Their expression only breaks down at Reynolds numbers exceeding approximately $3 \times \, 10^5$ when the type of flow around the sphere becomes turbulent. We have modeled this turbulent part of the flow as a step-function decrease of \cd to 0.2 \citep[e.g.,][]{Michaelides2006}. To illustrate the importance of including a varying drag constant, Fig. \ref{drag_evolution} shows the drag constant evolution of a 0.1 m \sio2 pebble impacting a 0.3 \me planet at escape velocity. During this impact, Kn is near unity at the edge of the planet and drops close to zero as the pebble approaches the core. In this regime, $\textrm{Kn} < 1$ and $\textrm{Ma} > 1$ for most of the impact, resulting in values of $C_\textrm{d}$ that are considerably different from the low-v drag expressions typically used in the planet formation literature \citep{Weidenschilling1977, Whipple1972}. \\

\subsection*{Ablation}\label{{Ablation}}
When an impactor enters the planet's envelope, it becomes surrounded by increasingly hot gas. We include two heating mechanisms, radiation and friction. Thermal radiation is approximated by perfect Planck curves. The net total of irradiated energy is then given by the Stefan--Boltzmann expression with the area of a sphere,
\begin{equation}\label{P_rad}
P_\textrm{rad} = 4 \pi R_\textrm{i}^2 \alpha_\textrm{i} \sigma_\textrm{sb} \left(T_\textrm{g}^4 - T_\textrm{s}^4\right)
,\end{equation}
where $\alpha_\textrm{i}$ is the pebble's average absorption constant, assumed to be unity. We use ambient gas temperatures even when incoming objects are supersonic. Gas temperatures and densities are highest near the core. We find that thermal ablation in combination with impactors being slowed down typically leads to very localized ablation in this inner region. 

For large or fast impactors, frictional heating becomes important. This is generally the dominant heating mechanism in the cold outer region of a planet's atmosphere where thermal ablation is minimal but the impactors travel at high velocities. These objects lose significant amounts of kinetic energy, a fraction of which ($f_\textrm{h}$) gets converted into heat:
\begin{equation}\label{P_fric}
P_\textrm{fric} = f_\textrm{h} F_\textrm{drag} v_\textrm{i}
\end{equation}
Estimating the numerical value of $f_\textrm{h}$ is a complex problem. Most works on impacts assume it to be equal to some constant of varying size \citep[e.g.,][]{Podolak1988, Pinhas2016}. We allow $f_\textrm{h}$ to vary during the impacts and evaluate $f_\textrm{h}$ based on gas characteristics with the expressions from \citet{Melosh2008},
\begin{equation}\label{fh}
f_\textrm{h} = \frac{8}{\gamma_\textrm{g}}\left(\frac{\textrm{Nu}}{\textrm{Re} \textrm{Pr}}\right)\frac{r'}{C_\textrm{d}}
,\end{equation}
where $r'$ is the recovery factor, evaluated as $r' = \left( \textrm{Pr}\right)^{\frac{1}{3}}$ \citep{Mills1999}. In addition to Re, $f_\textrm{h}$ also depends on the Nusselt (Nu) and Prandtl (Pr) numbers
\begin{equation}\label{Nusselt}
\textrm{Nu} = \frac{Nu_\textrm{c}}{1 + 3.42 M' Nu_\textrm{c} / \textrm{Re} \textrm{Pr}}
,\end{equation}
\begin{equation}\label{Prandtl}
 \textrm{Pr} = \frac{\eta_\textrm{g} C_\textrm{p,g}}{k_\textrm{g}}
,\end{equation}
where $C_\textrm{p, g}$ is the specific heat of the gas, $Nu_\textrm{c}$ and $M'$ are auxiliary functions
\begin{equation}\label{Nusselt_crit}
Nu_\textrm{c} = 2 + 0.459 \textrm{Re}^{0.55} r'
,\end{equation}
\begin{equation}\label{adapted Mach}
M' = \frac{\textrm{Ma}}{1 + 0.428 \textrm{Ma} \left(\gamma_\textrm{g} + 1\right) / \gamma_\textrm{g}}
,\end{equation}
and $\eta_\textrm{g}$ is the dynamic viscosity of the gas, the components $k$ of which we evaluate with the common expression from kinematic theory \citep[e.g.,][]{Dean1985}
\begin{equation}\label{adapted Mach2}
\eta_\textrm{k} = \frac{5}{16 d_\textrm{k}^2} \sqrt{\frac{m_\textrm{k} k_\textrm{b} T_\textrm{g}}{\pi}}
,\end{equation}
where $d_\textrm{k}$ and $m_\textrm{k}$ are the molecular diameter and weight of the gas constituent and $k_\textrm{b}$ is the Boltzmann constant. The values of $\eta_\textrm{k}$ are summed by volume fraction to yield $\eta_\textrm{g}$. In the simulated impacts of pebbles, the friction factor is found to vary between approximately 0.5 and 0.05. \\

As an impactor heats up, its inner region is initially shielded from the heat. This leads to the formation of a temperature gradient. The importance of this gradient can be inferred from the Biot number
\begin{equation}\label{Biot}
\textrm{Bi} = \frac{h' L}{k_\textrm{i}} 
,\end{equation}
where $h'$ is the characteristic heat transfer coefficient; $L$ is a characteristic length equal to $\frac{R}{3}$ for a sphere; and $k_\textrm{i}$ is the thermal conductivity, which we take equal to 2 $\rm {W} \, \rm{K} \, \rm{m}^{-1}$. This is a commonly used value, based on data from \citet{Powell1966} \citep[e.g.,][]{Adolfsson1996}. In the context of planetary entry, $h'$ can be approximated as \citep{Love1991}
\begin{equation}\label{Biot2}
h' = \sigma_\textrm{sb} \Delta T^3
,\end{equation}
where $\Delta T$ is the temperature difference between the local gas and the impactor's surface. Whenever the Biot number is less than 0.1, an object can roughly be considered as isothermal. A typical first order approach to modeling a temperature gradient is setting the Biot number equal to 0.1 and interpreting the resultant radius as an isothermal layer of temperature $T_\textrm{s}$ with depth and mass \citep[e.g.,][]{Love1991, Mcauliffe2006}
\begin{align}\label{Isothermal_layer}
d_\textrm{iso} &= \textrm{min}\left(0.3 \frac{k}{\sigma_\textrm{sb} \Delta T^3}, R_\textrm{i}\right) ,\\
M_\textrm{iso} &= \frac{4}{3} \pi r^3 \rho_\textrm{i} \left(R_\textrm{i}^3 - \left(R_\textrm{i} - d_\textrm{iso}\right)^3\right).
\end{align}
The total heat contents of the impactor are therefore approximated by $C_\textrm{p}\left(T_\textrm{s} M_\textrm{iso} + T_\textrm{disk}\left(M-M_\textrm{iso}\right)\right)$, where the temperature of the interior region is assumed to be equal to that of the disk, $T_\textrm{disk}$.
When the temperature of the surface layer has increased sufficiently, atoms start to evaporate into the envelope. This process can be characterized by the Langmuir formula
\begin{equation}\label{total power}
\dot{M_\textrm{i}} = 4 \pi R_\textrm{i}^2 P_\textrm{v} \sqrt{ \frac{m_\textrm{g}}{ 2 \pi k_\textrm{b} T_\textrm{s}}}
,\end{equation}
where $M_\textrm{i}$ is the impactor mass, $m_\textrm{g}$ is the mean molecular weight of the gas, and $P_\textrm{v}$ is the vapor pressure of the impactor (see Sect. \ref{High-Z enrichment and rainout}). Evaporation prevents the impactor from heating indefinitely. The surface temperature starts to decrease again when the energy loss through evaporation exceeds the incoming heat. Using the conservation of energy, the impactor's surface temperature can be obtained from
\begin{multline}\label{total power2}
 4 \pi R_\textrm{i}^2 \alpha_\textrm{i} \sigma_\textrm{sb} \left(T_\textrm{g}^4 - T_\textrm{s}^4\right) + f_\textrm{h} F_\textrm{drag} v_\textrm{i}- 4 \pi R_\textrm{i}^2 P_\textrm{v} \sqrt{ \frac{m_\textrm{g}}{ 2 \pi k_\textrm{b} T_\textrm{s}}} E_\textrm{vap} \\
 = C_\textrm{p} \left(M_\textrm{iso} \dot{T}_\textrm{s} +\dot{ M}_\textrm{iso} \left(T_\textrm{s} - T_\textrm{disk}\right)\right),
\end{multline}
where $E_\textrm{vap}$ is the latent heat upon evaporation, taken to be equal to $8.08 \times \, 10^6$ \Jkg1 for \sio2 as in \citet{Podolak2015}, and $C_\textrm{p}$ is the specific heat of the impactor. This is a temperature-dependent parameter that we evaluated with a fit reported on the Chemistry Webbook of the National Institute of Standards and Technology (NIST), based on original measurements by \citet{Chase1998} (see Eq. 25).

In summary, the terms in Eq. (\ref{total power}) denote contributions respectively from radiative cooling, frictional heating, evaporative cooling, and internal energy. This temperature balance is invoked to yield the equilibrium temperature of the isothermal layer $T_\textrm{s}$ with a resolution of $10^{-3}$ K at any time step. Such a precise temperature determination is necessary to ensure numerical stability despite the strong temperature dependence of the vapor pressure.

\subsection{Planetary structure}\label{Planetary structure}
We integrate the planet's atmosphere from the outside in, starting at either the Hill or the Bondi radius, depending on which is smaller. At this radius the envelope's composition and conditions are taken to be the same as the disk's. Our code solves for two boundary conditions at the core surface: the mass condition $M\left(r = r_\textrm{core}\right) = M_\textrm{core}$ and the luminosity condition
\begin{equation}\label{impact energy}
L_\textrm{core} = \chi_\textrm{i, core} \frac{\dot{M}_\textrm{acc} v_\textrm{i, core}^2}{2} 
,\end{equation}
where $\chi_\textrm{i, core}$ and $v_\textrm{i, core}$ are the surviving mass fraction and velocity of impactors that reach the core and $\dot{M}_\textrm{acc}$ is the solids accretion rate. We consider accretion to be the only energy source at the core and neglect any additional contributions, such as radioactive heating or core contraction. The total luminosity and envelope mass are adjusted such that these conditions are matched at every time step. Disk gas flows into the envelope to fit the increased interior mass. 

The structure of the planet is divided into two parts: a solid core and a gaseous envelope. We use a typical constant density for the core of 3.2 $\times \, 10^3$ \kgm3 \citep[e.g.,][]{Pollack1996, Hubickyj2005}. The core grows only when solid material reaches its boundary. In our model, this can occur in two ways. Either by direct impacts of accreting material or by the rainout of ablated silicates when the vapor saturation of the envelope is exceeded. The envelope's structure is calculated with the one-dimensional spherically symmetric stellar structure equations
\begin{equation}
\frac{dP}{dr} = \frac{- G M \rho}{r^2}
,\end{equation}
\begin{equation}
\frac{dM}{dr} = 4 \pi \rho r^2
,\end{equation}
\begin{equation}
\frac{dT}{dr} = \frac{dP}{dr}\frac{T}{P} \nabla_\textrm{th},\\
\end{equation}
where $G$ is the gravitational constant, $P$ is pressure, $T$ the temperature, $M$ the total mass interior to radius $r$, $\rho$ the envelope density, and $r$ the distance to the planet's center. We use the Schwarzschild criterion to determine the thermal gradient $\nabla_\textrm{th}$ as $\nabla_\textrm{th} = \textrm{min}\left(\nabla_\textrm{conv}, \nabla_\textrm{rad}\right)$, where $\nabla_\textrm{conv}$ is approximated by the adiabatic gradient $\nabla_\textrm{ad}$, which we take as a constant based on the local gas composition
\begin{equation}
\nabla_\textrm{ad} = \frac{\gamma_\textrm{g} - 1}{\gamma_\textrm{g}},\\
\end{equation}
where $\gamma_\textrm{g}$ is the average adiabatic index of the gas, defined as the ratio of the specific heat under constant pressure and volume $C_\textrm{p, g} / C_\textrm{v, g}$. We use the fitting formula from \citet{Chase1998} to evaluate the components $C_\textrm{p, k}$ of $C_\textrm{p, g}$ as a function of temperature
\begin{equation}
C_\textrm{p, k} = A_\textrm{k} + B_\textrm{k}T_\textrm{g} + C_\textrm{k}T_\textrm{g}^2 + D_\textrm{k}T_\textrm{g}^3 + E_\textrm{k} / T_\textrm{g}^2
,\end{equation}
\begin{equation}
C_\textrm{p, g} = \sum_\textrm{k} \frac{C_\textrm{p, k}}{\mu_\textrm{k}}
,\end{equation}
where $A_\textrm{k} - E_\textrm{k}$ are constants corresponding to the component k (here $k = $ (H$_2$, He, SiO$_2$, H$_2$O)). To calculate the components $C_\textrm{v, k} = C_\textrm{p, k} / \gamma_\textrm{k}$, we assume constant adiabatic indices $\gamma_\textrm{k}$ equal to $\gamma_{\textrm{H}_2} = 1.4$, $\gamma_\textrm{He} = \frac{5}{3}$, and $\gamma_{\textrm{SiO}_2} = 1.2$ to find
\begin{equation}
C_\textrm{v, g} = \sum_\textrm{k} \frac{f_\textrm{g, k}}{\gamma_\textrm{k}} \frac{C_\textrm{p, k}}{\mu_\textrm{k}}
,\end{equation}
where $f_\textrm{g, k}$ are the local gas mass fractions. For the radiative outer zone of the envelope, we use
\begin{equation}
\nabla_\textrm{rad} = \frac{3 \kappa L P}{64 \pi \sigma_\textrm{sb} G M T^4},\\
\end{equation}
where $\kappa$ is the opacity and $L$ the luminosity. Following \citet{Ormel2014}, the opacity is assumed to originate only from the gas and is approximated by the analytical expression from \citet{Bell1994},

\begin{equation}
    \label{eq:BellLin}
\kappa = 10^{-9} \rho^{\frac{2}{3}} T^3, \\
\end{equation}
valid for a solar-type composition. In the inner, highly enriched regions \eq{BellLin} would no longer apply, but these regions are adiabatic and thus $\kappa$ does not enter the structure equations. We do not assume a grain opacity, as these are likely to be very small as a result of coagulation and sedimentation processes \citep{Movshovitz2010,Mordasini2014,Ormel2014}.

The luminosity has four components, kinetic energy deposition by the impactors, (negative) latent heat release upon ablation, high-Z rainout, and contraction, 
\begin{equation}
\frac{dL}{dr} = \left(\frac{d K_\textrm{i}}{dr} - E_\textrm{vap}\frac{d \chi_\textrm{i}}{dr}\right) \dot{M}_\textrm{acc} + \frac{G M \dot{M}_\textrm{rain}}{r^2} - 4 \pi r^2 \rho P \dot{V} 
,\end{equation}
where $K_\textrm{i}$ is the kinetic energy of an impactor, $\dot{M}_\textrm{rain}$ is the local rainout rate, and $V$ is the specific gas volume $\frac{1}{\rho}$. The core luminosity is determined by the mass fraction ($\chi_\textrm{i, core}$) of the pebbles that reaches the core and their kinetic energy (see Eq. (\ref{impact energy})). The structure equations are supplemented by the ideal gas equation of state for a mixed composition:
\begin{equation}
\rho = \frac{P m_\textrm{g}}{k_\textrm{b} T}
.\end{equation}
The range of conditions we experience in our simulations is still well approximated by the ideal gas equation (see Sect. \ref{Model sensitivities}). Temperatures inside the planet's inner region can exceed those required for the dissociation of hydrogen \citep{LeeEtal2014}, but these regions are modeled as silicate-dominated and contain little hydrogen. The effective difference with using a more sophisticated equation of state is therefore minor. An explicit test of this is included in Sect. \ref{Model sensitivities}.

The structure equations are integrated using a fourth-order Runge--Kutta scheme. We have set the time step $\Delta t = \Delta M / \dot{M}_\textrm{acc}$ such that the planet accretes 0.005 \me at every interval. The grid consists of set logarithmic distances to provide enough resolution near the core and consists of approximately 20,000 layers per integration. The reason for this high number is to ensure a smooth temperature profile of the inner region of the planet. This reduces the number of iterations needed in the impact code and speeds up run-times.

\subsection{High-Z enrichment and rainout}\label{High-Z enrichment and rainout}
We consider the planet to be embedded in a gaseous disk with mass fractions of 75\% molecular hydrogen and 25\% helium. The initial envelope and the accreted gas do not contain any high-Z materials. However, as the protoplanet is impacted by pebbles, some of their material ablates into the atmosphere before it reaches the core. We consider the effects this has on the planet's evolution and subsequent impacts by locally increasing the mean molecular weight where high-Z material is present in vapor form. The maximum fraction (by mass) of high-Z vapor that the gas can contain depends on the local silicate vapor and total pressure
\begin{equation}\label{vapor fraction}
f_{\textrm{g}, \textrm{SiO}_2, \textrm{max}} = \frac{\mu_{\textrm{SiO}_2}}{\mu_\textrm{g}} \frac{P_\textrm{vap}^{\textrm{SiO}_2}}{P}
,\end{equation}
where $f_{\textrm{g}, \textrm{SiO}_2}$ is the local mass fraction of silicates in vapor phase. Equation (\ref{vapor fraction}) can be interpreted as the high-Z saturation curve of the gas, depending on pressure and temperature. Vapor pressures can be expressed by the Clausius--Clapeyron equation 
\begin{equation}\label{vapor pressure}
P_\textrm{vap}^{\textrm{SiO}_2} = \textrm{exp}[a_0 - \frac{a_1}{T + a_2}]
,\end{equation}
where $a_0 = 29.5, a_1 = 46071.4\, \textrm{K}$, and $a_2 = 58.9\, \textrm{K}$ are constants determined by the fit published on NIST, based on original data from \citet{Stull1947}. We iterate on $\mu_\textrm{g}$ at every time step to ensure an accurate estimation. When the gas temperature increases such that $f_{\textrm{g}, \textrm{SiO}_2} > 1$, we limit $f_{\textrm{g}, \textrm{SiO}_2}$ to unity. In this case, the atmospheric layer consists entirely of silicate gas. The temperature dependency of the vapor pressure limits the presence of silicate vapor to the envelope's warmer interior layers. Finding the partial pressure also requires identifying how much total silicate mass is present in a given layer. This is affected by the amount of material that has been ablated up to that point, but also depends on what happens to the silicates after ablation (i.e., outward mixing or settling). To simplify matters, we look at two contrasting scenarios:
\begin{enumerate}
\item
In the mixing case, we assume that all the ablated material mixes uniformly through the entire envelope. This includes the material in solid and in vapor form. We track the total high-Z mass that has ablated up to that point in time, $M_\textrm{Z, abl}$. Meanwhile, the planet has also attracted hydrogen and helium gas masses $M_{\textrm{H}_2}$ and $M_\textrm{He}$ from the disk. Complete mixing then leads to one global silicate mass fraction (vapor + solids) that is calculated as
\begin{equation}\label{mixing silicates}
f_{\textrm{SiO}_2} = \frac{M_\textrm{Z, abl}}{M_\textrm{Z, abl} + M_{\textrm{H}_2} + M_\textrm{He}}
\end{equation}
In the outer layers, $f_{\textrm{SiO}_2} > f_{\textrm{g}, \textrm{SiO}_2}$ and most of the silicate mass is in solid form. This trend continues inward up to the point that the temperature of a layer is sufficiently high that $f_{\textrm{SiO}_2} = f_{\textrm{g}, \textrm{SiO}_2}$. Here, mixing limits the silicate vapor mass fraction to $f_{\textrm{SiO}_2}$ of Eq. (\ref{mixing silicates}). Direct core growth in this scenario only continues as long as some of the impactors can directly reach the core.
\item
In the rainout case, any material in excess of what can be contained in vapor (see Eq. (\ref{vapor fraction})) falls to the layer below. The resulting rainout of high-Z material increases the planet's luminosity as descending silicate mass loses gravitational energy on its way to the core. If the envelope cannot absorb all the ablated material, the rest of it adds to the core mass. This allows the core to grow beyond the point that impactors fail to reach the core directly. Direct core growth ceases when all the accreting solids can be fully absorbed by the envelope as vapor.
\end{enumerate}

Vapor pressures rise quickly as the local gas temperature increases. Close to the core, conditions can be such that the vapor pressure exceeds the total pressure, and the high-Z mass fraction approaches one. The high mean molecular weight of this high-Z layer surrounding the core leads to very steep pressure and density curves, as can be seen in Fig. \ref{interior_profiles}.

\section{Results}\label{Results}

\subsection{Standard model}\label{Standard model}

\begin{table}
\caption{Descriptions and values of the standard model parameters}
\label{table_parameters}
\centering
\begin{tabular}{l l l}
\hline\hline
Parameter & Description & Value \\ 
\hline                  
$R_\textrm{i}$ & Impactor radius (m)  & 0.1 \\      
$f_{\textrm{H}_2}, f_\textrm{He}  $   & Disk gas mass fractions & 0.75, 0.25 \\
$T_\textrm{disk}$        & Local disk temperature (K)   & 150   \\
$d_\textrm{planet}$                          & Orbital radius (AU)          & 5.2 \\ 
$M_\star$      & Mass of the central star ($\textrm{M}_\odot$) & 1  \\
$\rho_\textrm{disk}$     & Local disk density (\kgm3)   & $5 \times \, 10^{-8}$               \\
$\dot{M}_\textrm{acc}   $      & Solids accretion rate ($\textrm{M}_\oplus \, \textrm{yr}^{-1}$)   & $10^{-5}$                       \\ 
$\mu_\textrm{Z}   $      &  Molecular weight \sio2 ($\textrm{g mol}^{-1}$)   & 60.08                \\ 
$\gamma_\textrm{Z}   $      &  Adiabatic index \sio2   & 1.2                      \\ 
\hline                                   
\end{tabular}
\end{table}

\begin{figure*}[h!]
\resizebox{\hsize}{!}{\includegraphics{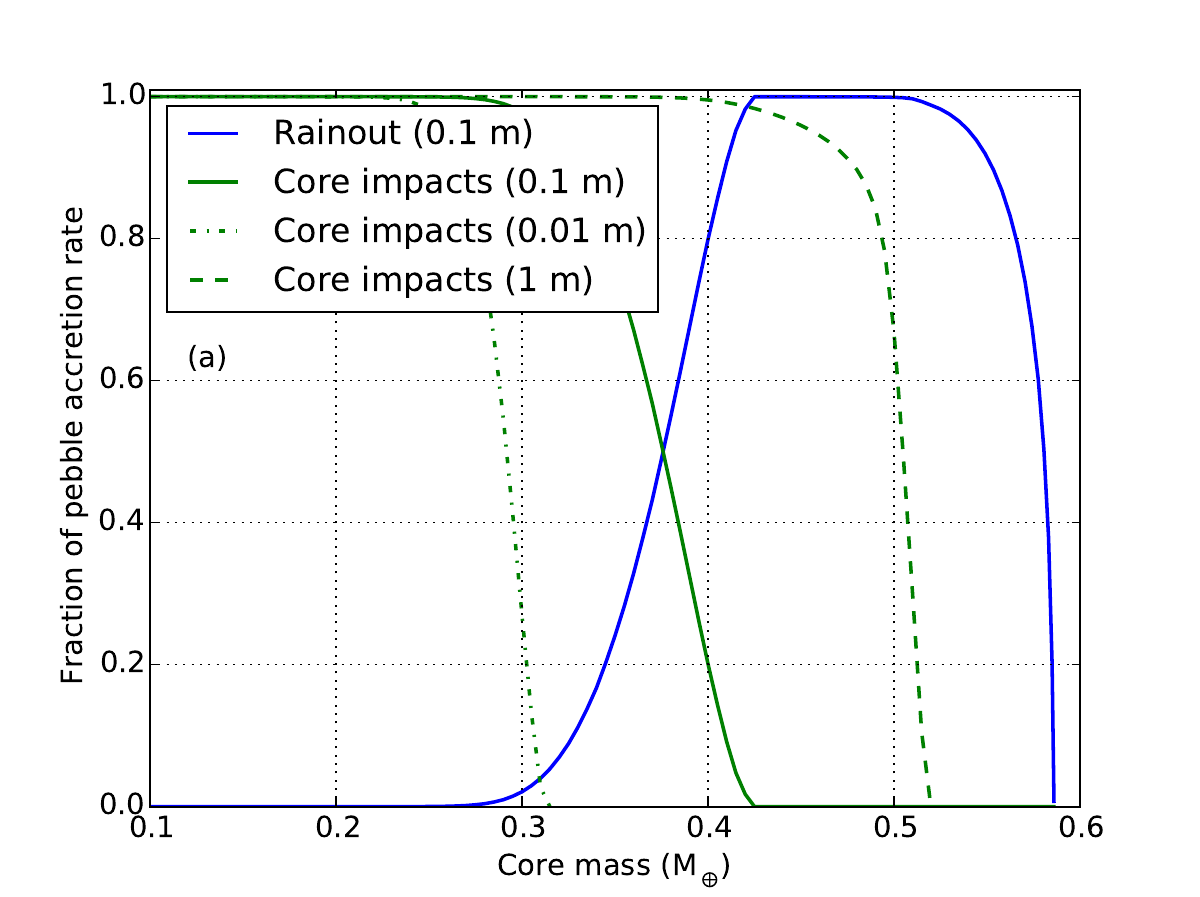} \includegraphics{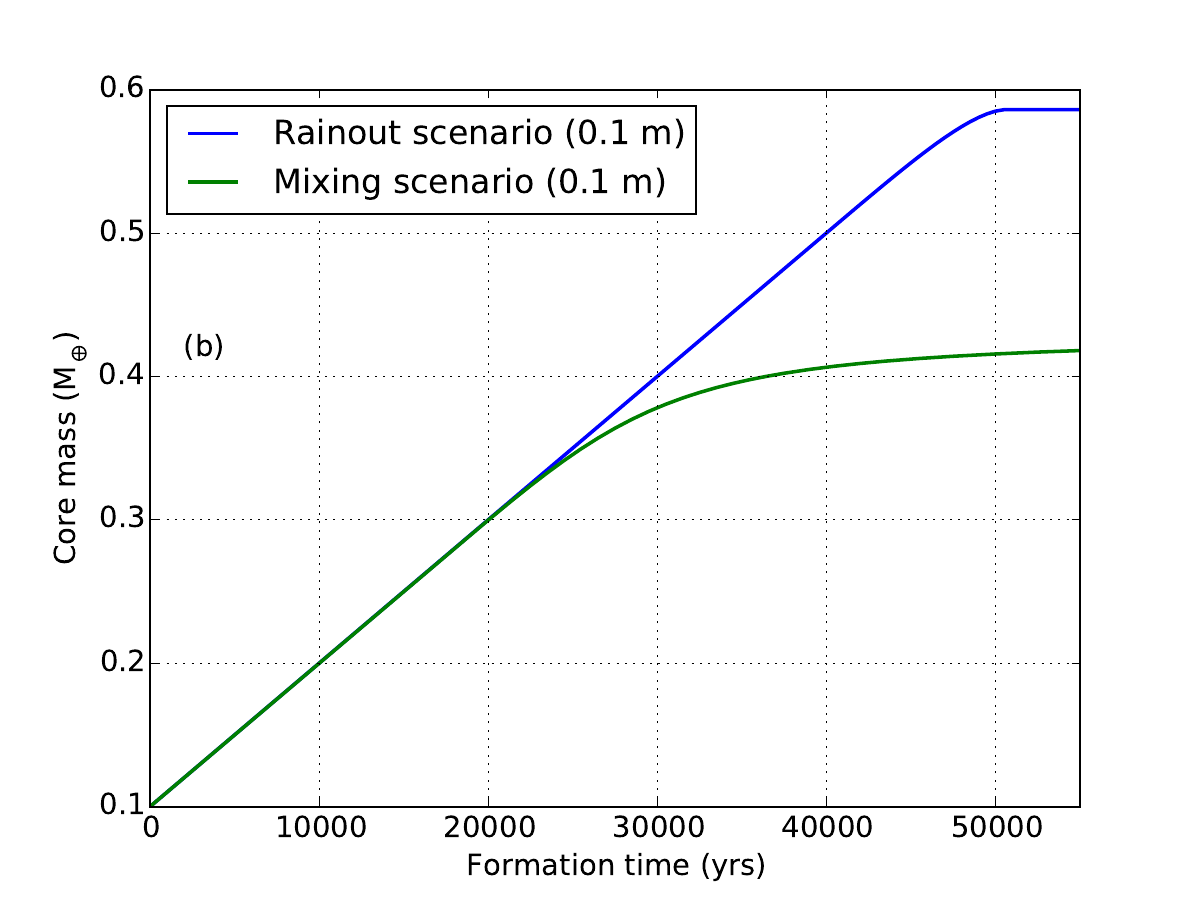}} 
\caption{Core growth by pebble accretion. Left: Panel (a) shows the mass fractions of solids impacting the core directly (green curves) or reaching the core by rainout of ablated material (blue curve). The differently spaced green curves correspond to three sizes of pebbles: 0.1 m (solid), 1 m (dashed), 0.01 m (dash-dotted). Right: Panel (b) shows the growth of the core over time. In this figure, the rainout scenario is indicated by the blue curve and the mixing case is shown by the green curve. Both curves in panel (b) correspond to 0.1 m impactors. \label{core_growth_mechanisms}}
\end{figure*}

The simulations shown here use the parameters from Table \ref{table_parameters} unless specifically stated otherwise. We simulate the planet at a distance of 5.2 AU from the central star, embedded in a disk with a typical local temperature and density of 150 K and $5 \times \, 10^{-8}$ \kgm3 \citep{Hubickyj2005}. We use a standard pebble accretion rate of $10^{-5}$ $\textrm{M}_\oplus \, \textrm{yr}^{-1}$ \citep{Lambrechts2014}. The goal is to estimate to what mass a core can grow before direct core accretion stops. This is either when impacting pebbles fully evaporate in our mixing scenario, or when the atmosphere can absorb all ablated material in the rainout case. Figure \ref{core_growth_mechanisms} shows the result of our \sio2 pebble simulations for both of these assumptions.

In general, we can distinguish three phases of core growth:
\begin{enumerate}
\item
When the core mass is still below $\sim$ 0.23--0.39 \me, depending on impactor size, all pebbles can reach the core without experiencing significant thermal ablation. 
\item
After this point, ablation becomes increasingly severe and a decreasing mass fraction impacts the core directly. The planet's temperature and pressure are still too low for the envelope to be able to retain a significant amount of silicate vapor, causing the rest to rain out. If the ablated material mixes outwards instead, core growth slows down.
\item
At around 0.50 \me, the planet's envelope mass and temperature have increased sufficiently for absorption of high-Z vapor in the envelope's inner region to become significant. Impactors can be fully ablated in the atmosphere and direct core impacts terminate. The planet becomes heavily enriched during this third phase, leading to the formation of a high-Z layer around the core with a very high density and temperature.
\end{enumerate}

\begin{figure}[h!] 
\centering
\includegraphics[width=\hsize]{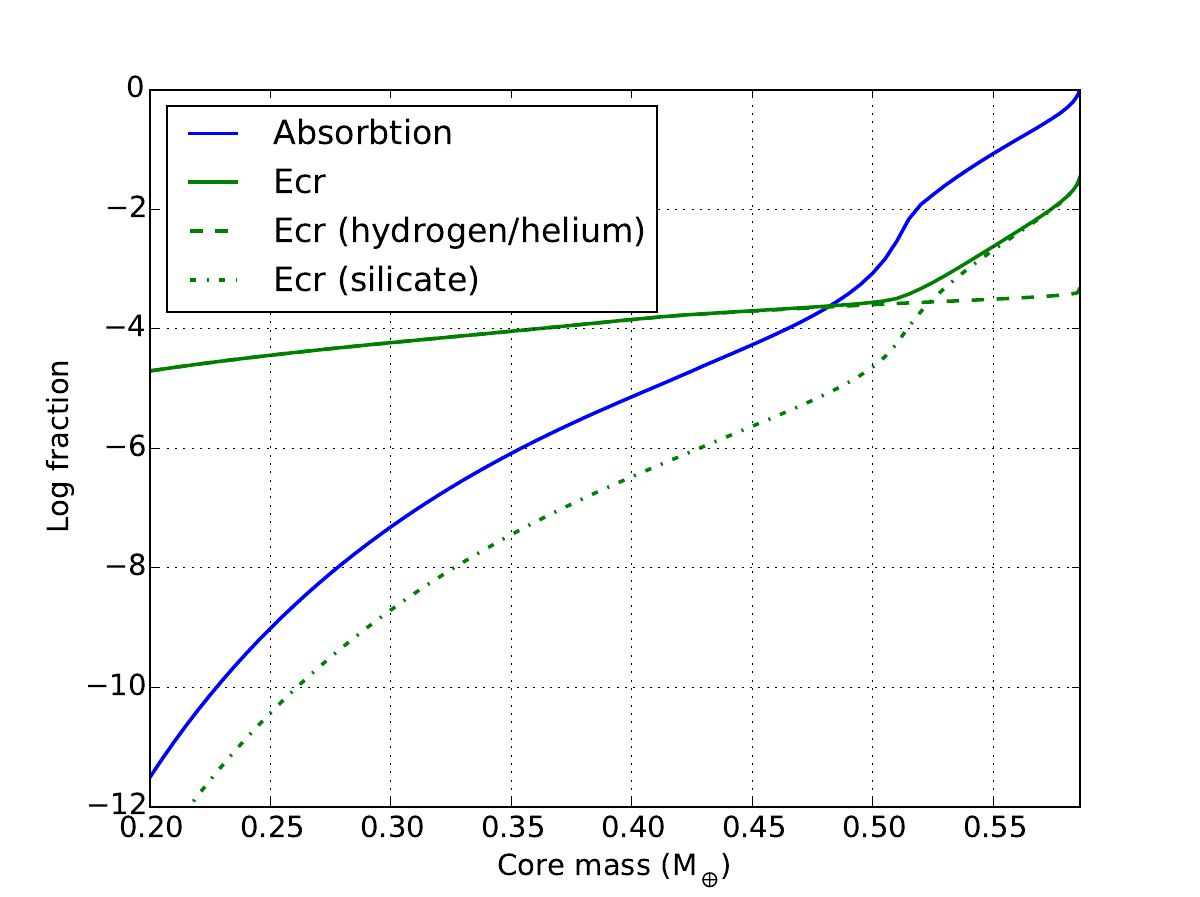} 
\caption{ Absorption fraction (blue) and envelope-to-core ratio ($ecr$; green) of a planet impacted by 0.1 m pebbles. The green curves indicate the $ecr$, defined as $M_\textrm{env} / M_\textrm{core}$. The dash-dotted curve is its silicate mass fraction and the dashed curve is its hydrogen and helium mass fraction. They sum to the total $ecr$, shown by the solid green line. The blue absorption curve indicates the mass fraction of the accreting solids that can be contained by the envelope as vapor.\label{abs_ecr}}
\end{figure}

To further illustrate the core growth process, we have plotted the envelope-to-core ratio of a growing planet along with its absorption rates in Fig. \ref{abs_ecr}. The absorption rate is defined as the fraction of ablated accreting material that stays in the atmosphere in vapor form. The planet's envelope-to-core ratio starts to increase rapidly after its core mass reaches $\sim$ 0.5 \me. This coincides with the formation of a high-Z layer near the core, as can be seen by the simultaneous increase in the absorption rate. The envelope mass is at first mostly due to hydrogen and helium gasses. When absorption becomes significant, the envelope's metal rich inner region becomes very hot and dense and its constituting silicate vapor starts to dominate the envelope mass.

\begin{figure*}[h!]
\resizebox{\hsize}{!}{\includegraphics{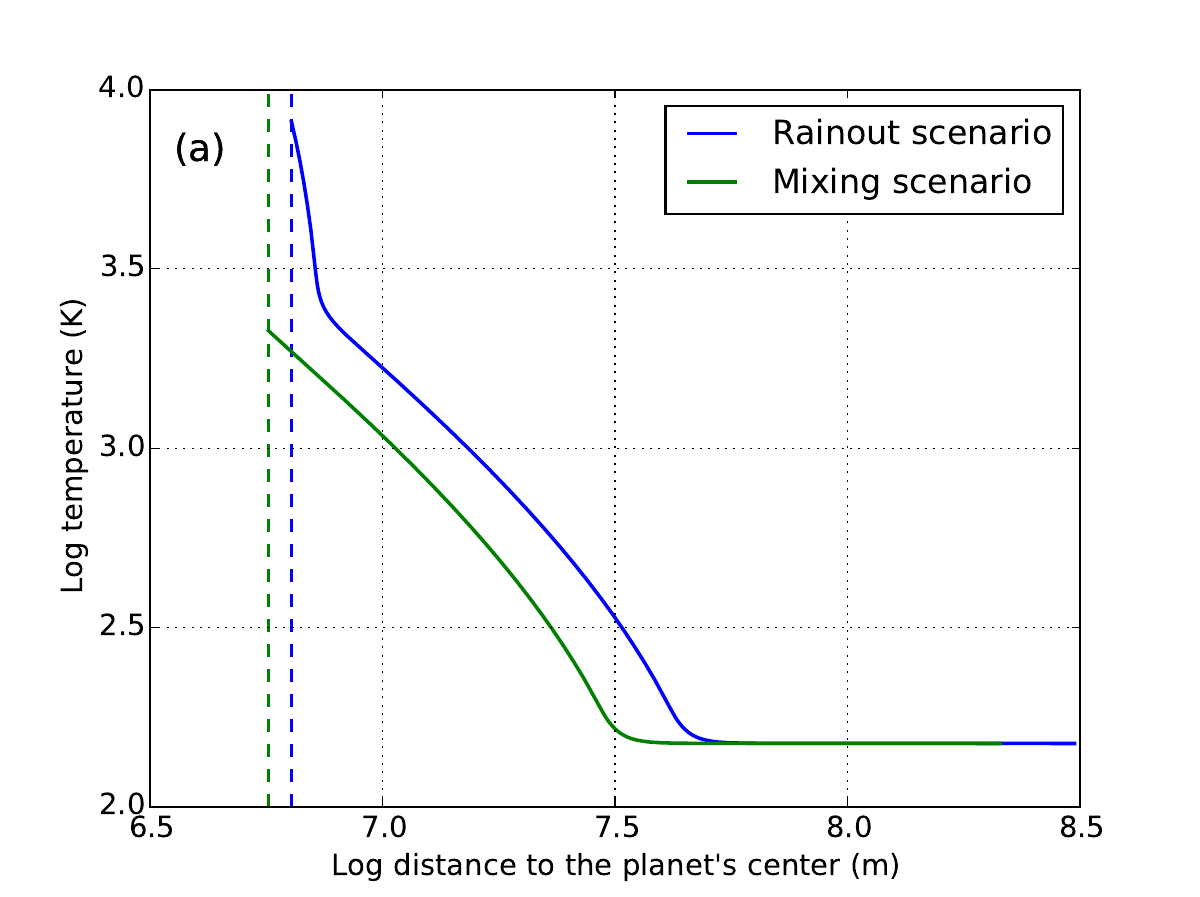} \includegraphics{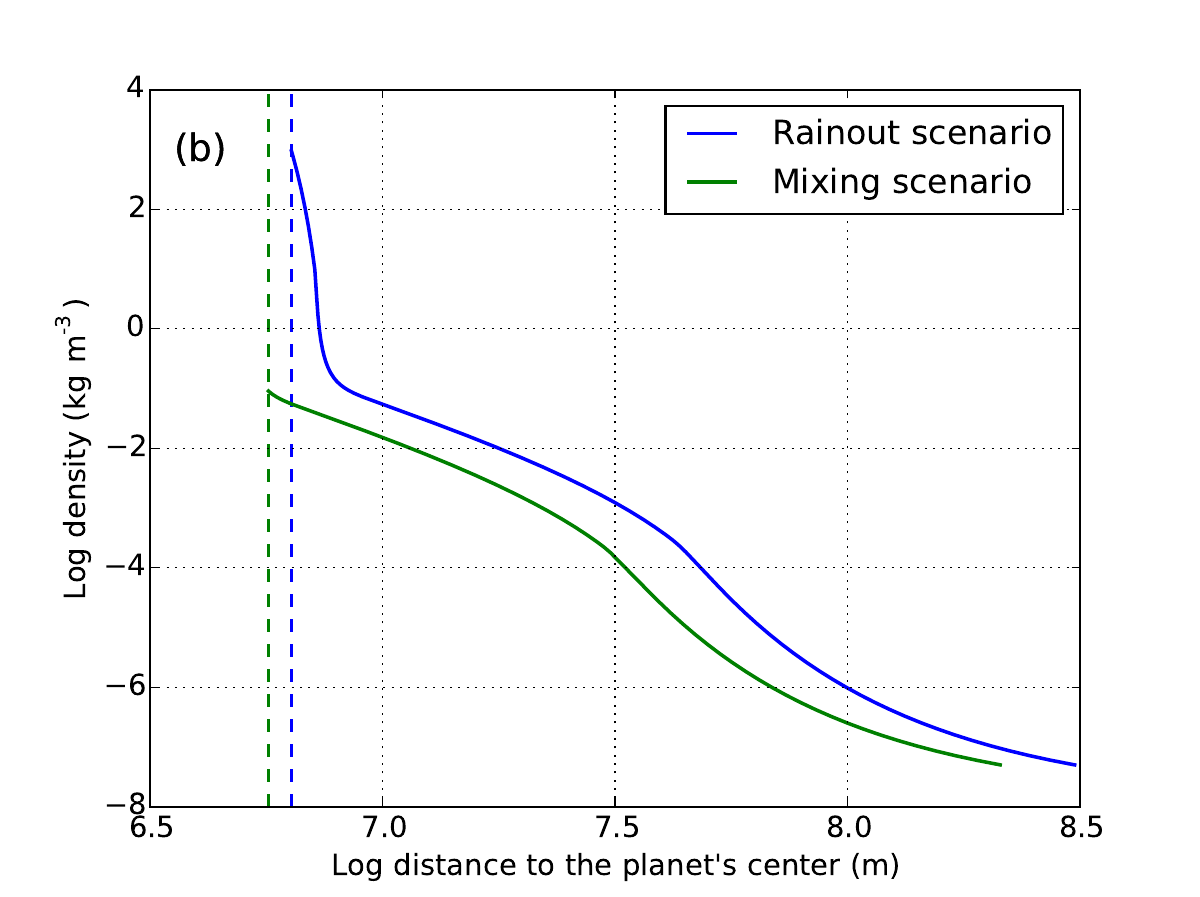}} 
\resizebox{\hsize}{!}{\includegraphics{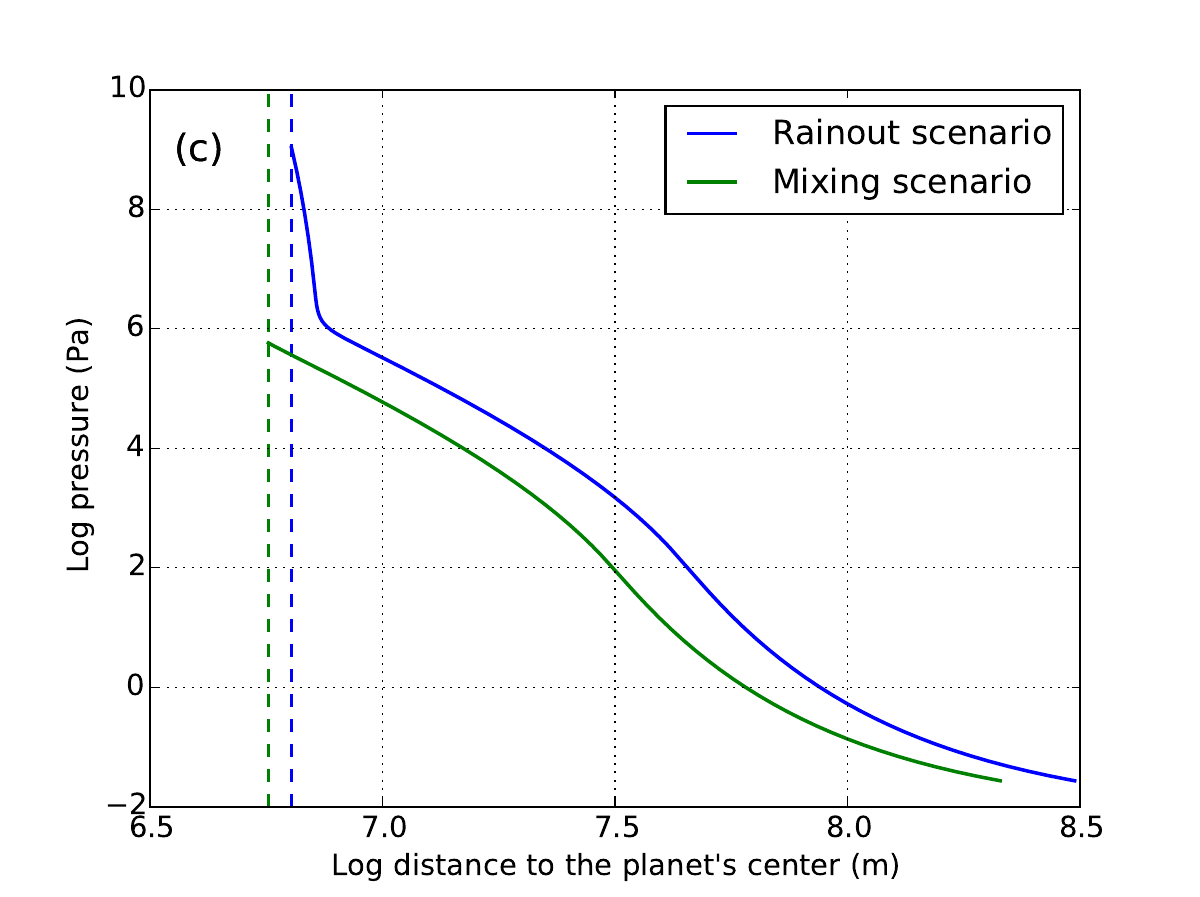} \includegraphics{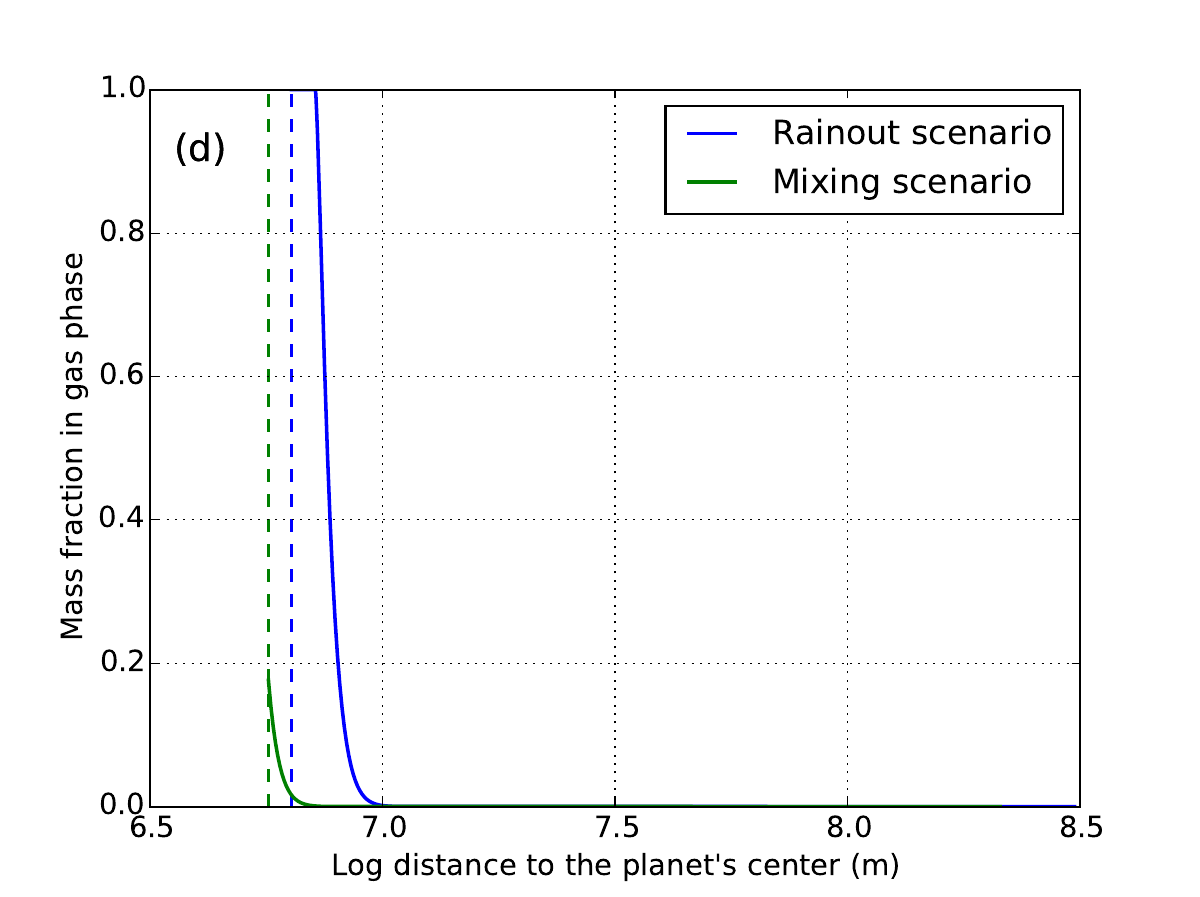}} 
\caption{Temperature (a), density (b), pressure (c), and high-Z enrichment (d) interior curves of a planet formed by pebble accretion. They are plotted at the point in the planet's evolution when core growth is equal to 10\% of the pebble accretion rate. This corresponds to core masses of 0.58 \me in the rainout scenario (blue curves) and 0.41 \me in the mixing case (green curves). The dotted vertical lines indicate the position of the core.}
\label{interior_profiles}
\end{figure*}

The temperature, density, pressure, and high-Z enrichment curves of the resulting planet are plotted in Fig. \ref{interior_profiles}. The gas of the enriched inner region has a much higher mean molecular weight in the rainout case, causing it to become very hot and dense.

\subsection{Model sensitivities and limitations}\label{Model sensitivities}
To investigate the sensitivity of our results to the variation of several parameters, we performed the same simulations with different pebble sizes, disk orbital radii, solid accretion rates, or $\textrm{SiO}_2$ characteristics (see Table \ref{parameter study}). The disk conditions for different planet positions correspond to an extrapolation of the conditions at 5.2 AU. We used MMSN scaling, so $T_\textrm{disk}$ $\alpha$ $r^{-0.5}$ and $\rho_\textrm{disk}$ $\alpha$ $r^{-2.75}$ \citep[e.g.,][]{Weidenschilling1977, Hayashi1981}. The last columns of Table \ref{parameter study} show the final core masses for various sets of parameters. These represent the points at which pebbles are either fully evaporated (mixing) or all ablated material can be absorbed as vapor (rainout). 

\begin{table}
\caption{ Variation in final core masses for different pebble sizes, disk conditions, and accretion rates. The final column (left and right) shows the core masses at which either pebbles are fully evaporated (mixing scenario; left) or all ablated material can be absorbed by the envelope as vapor (rainout scenario; right). The standard model parameters are given in Table \ref{table_parameters}.}
\label{parameter study}
\centering
\begin{tabular}{l l l c}
\hline\hline  
Parameter &  Variation & \multicolumn{2}{l}{Final core masses (\me)} \\    
\hline    
--   & --  & 0.42 & 0.59  \\              
$R_\textrm{i}$   & 0.01 (m)   & 0.31 & 0.59 \\      
$R_\textrm{i}$    & 1  (m)  & 0.52 & 0.58 \\
EOS   & QEOS (1) & 0.42 & 0.59 \\
$\dot{M}_\textrm{acc}$   & $10^{-4} \, (\textrm{M}_\oplus \, \textrm{yr}^{-1})$      & 0.41 & 0.55  \\
$\dot{M}_\textrm{acc}$   & $10^{-6} \,  (\textrm{M}_\oplus \, \textrm{yr}^{-1})$      & 0.42 & 0.65 \\ 
$d_\textrm{planet}$   & 10 (AU)   & 0.40 & 0.60 \\ 
$d_\textrm{planet}$   & 1 (AU)   & 0.41 & 0.54 \\ 
$d_\textrm{planet}$   & 0.5 (AU)   & 0.33 & 0.48 \\ 
$d_\textrm{planet}$   & 0.2 (AU)   & 0.23 & 0.37  \\ 
$\mu_{\textrm{SiO}_2}$   & 40 ($\textrm{g} \, \textrm{mol}^{-1}$) & 0.42 & 0.66  \\ 
$\mu_{\textrm{SiO}_2}$   & 18 ($\textrm{g} \, \textrm{mol}^{-1}$) &  0.43 & 0.97  \\ 
$\mu_{\textrm{SiO}_2}$   & 10 ($\textrm{g} \, \textrm{mol}^{-1}$) &  0.44 & 1.49 \\ 
$\gamma_{\textrm{SiO}_2}$   & 1.4 & 0.42 & 0.66  \\ 
$\gamma_{\textrm{SiO}_2}$   & 1.1 &  0.42 & 0.57 \\ 
\hline                                   
\end{tabular}
\tablebib{
(1)~\citet{Vazan2013}.
}
\end{table}

To zeroth order, the computed final core masses shown in Table \ref{parameter study} can be predicted by the point at which the core surface temperature first exceeds about 1600 K, and \sio2 starts evaporating. Our general picture of limited core growth does not change if we vary the pebble size, disk conditions, or the solids accretion rate. The inner atmosphere always reaches 1600 K at a core mass of $\sim$ 0.3 \me. As expected, smaller pebbles evaporate sooner than larger ones, thus leading to a lower core mass in the mixing case. Final core masses after rainout terminates, are less significantly influenced by pebble size because the envelope's absorption rates stay the same. 

These absorption rates do change when the accretion rate is varied, but then final core masses in the mixing and rainout cases do not differ much from 0.4 and 0.6 \me, respectively. This can be explained by noting that increasing the accretion rate of pebbles causes two competing effects on pebble impacts and absorption rates. Firstly, higher accretion rates lead to increased luminosity and thus to higher temperatures, causing more ablation. At the same time however, these higher temperatures reduce gas densities and thus inhibit gas drag. The result is that pebbles fully evaporate at a similar core mass, but at a significantly lower envelope mass (not shown in table). Similarly for absorption rates, higher accretion rates mean that the temperature at which silicates vaporize starts further from the core, but at a lower density. The same effects also influence the final core mass as a function of position, where a planet's envelope closer to the central star is hotter and less dense. In our model, final core masses tend to decrease as the planet's location moves closer to the star.

We find that the core formation process is most sensitive to changes in the mean molecular weight of the high-Z constituent \sio2, which determines the density scaling of the high-Z layer near the core and therefore has a great influence on how much vapor this layer can contain. Variation in this parameter is possible, as the pressure and temperature conditions near the core may allow for \sio2 dissociation \citep[e.g.,][]{Medvedev2016}, especially in the presence of hydrogen \citep[e.g.,][]{Elsayed2015, Soubiran2017}. We vary the mean molecular weight of the high-Z component all the way down to $\mu_{\textrm{SiO}_2} = 10 \textrm{g mol}^{-1}$. Our simulations show that in this case, the final core mass of the rainout case increases to 1.49 \me. Other equation-of-state effects are not found to be as significant. Table \ref{parameter study} shows that final core masses do not change when we switch from an ideal gas to a more sophisticated equation of state for a mixture of hydrogen, helium, and \sio2 of \citet{Vazan2013} that does not include \sio2 dissociation. We explain this with the observation that, while the temperatures and densities near the core can exceed those required for non-ideal gas effects such as hydrogen dissociation, this hot inner region consists mainly of silicates (see Fig. \ref{interior_profiles}). We have also included a variation in the adiabatic index because this value is less well known. It is found not to have a large impact on the resultant core masses. 

Finally, we performed an additional simulation without considering any enrichment effects, where all ablated material falls to the core. In this case we found that the core grew to 5.1 \me, at which point gas accretion became dynamically unstable and our simulation was stopped. This value for the critical core mass is consistent with previous calculations that assumed low (grain-free) opacities \citep[e.g.,][]{Mizuno1980, Hori2010}.

\subsection{Icy pebbles}\label{Icy pebbles}
In reality, pebbles outside the iceline are expected to be supplemented by \h2o ices \citep[e.g.,][]{ Schoonenberg2017, Ida2016b}. The impactor's composition is a very important parameter in our core formation model. It determines how rapidly the impactors ablate, how much of the ablated material can be absorbed by the envelope, and what the envelope's interior region looks like. As a comparison, we repeated the previously described simulations with entirely icy \h2o pebbles. We find that these icy pebbles can be fully ablated very early in the planet's evolution, before the planet's core has even grown to 0.05 \me. This is the case even if we simulate a planet that forms well beyond the \h2o iceline at 10 AU at a local disk temperature of 108 K. The subsequent absorption of the water vapor is also more efficient than the absorption of \sio2 vapor. The main determinant of these processes is the vapor pressure, which for water already becomes significant at temperatures of around 200 K, compared to 1600 K for \sio2. This leads to the envelope becoming dominated by water vapor at a low mass \citep{Venturini2015}. If we assume that the inner region of the envelope stays in vapor form, we find that direct core growth terminates before the core reaches 0.1 \me. However, we observe that conditions are such in the inner region that the water vapor near the core will liquify, leading to a more complex evolution of the planet, similar to that described by \citet{Chambers2017}.

\section{Comparison with planetesimal impacts}\label{planetesimal comparison}

\subsection{Mass deposition}\label{Mass deposition}

\begin{figure*}[h!]
\resizebox{\hsize}{!}{\includegraphics{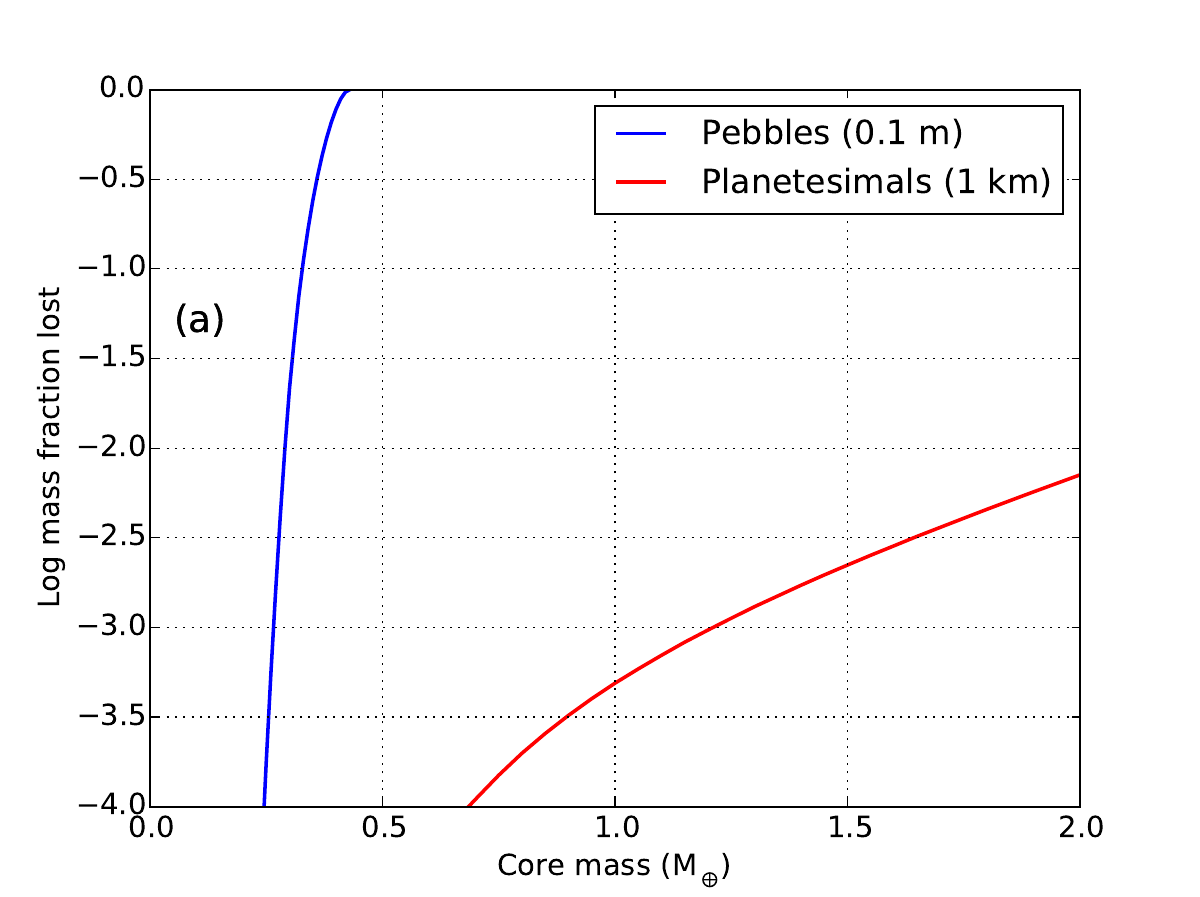} \includegraphics{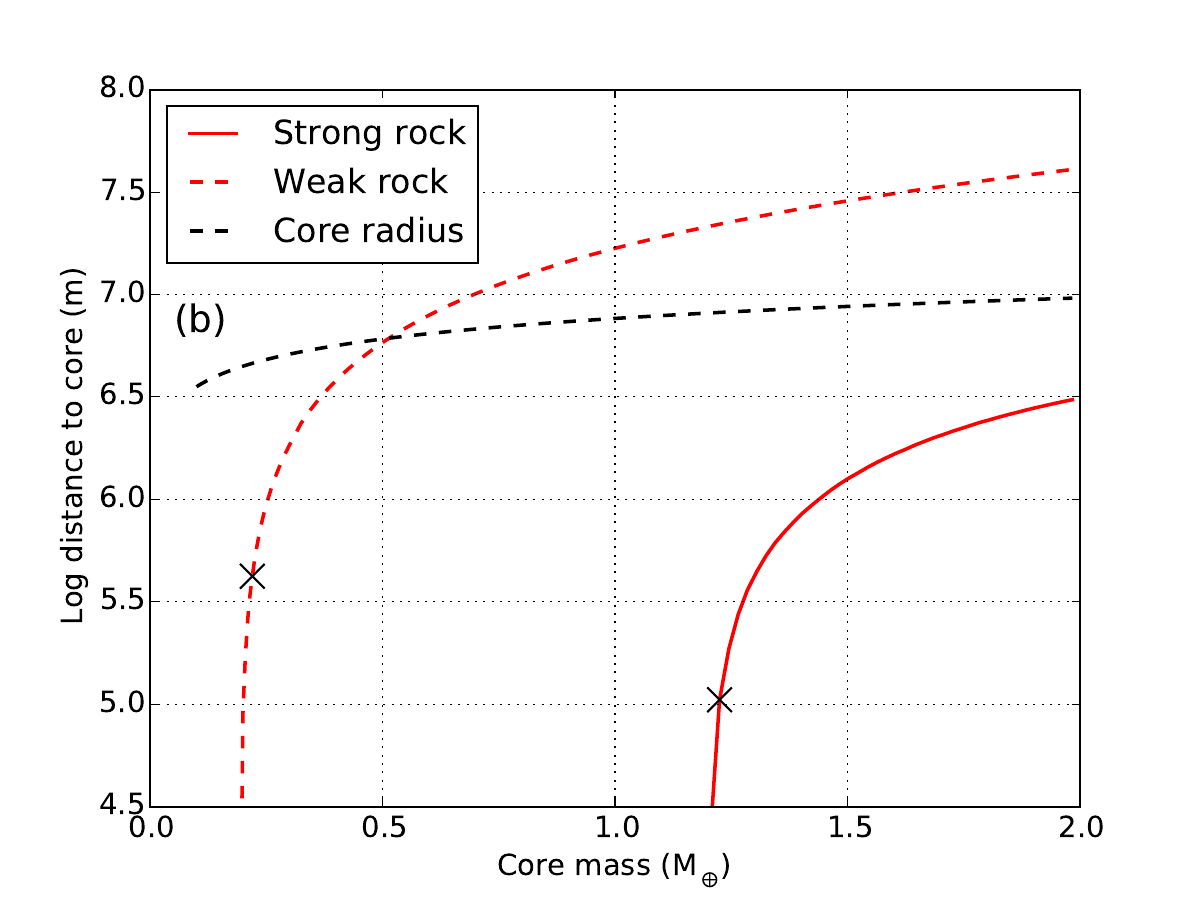}} 
\caption{
Mass deposition by planetesimals. Left: Panel (a) shows ablation curves for 1 km planetesimals (red line) and 0.1 m pebbles (blue line) impacting a growing protoplanet, ignoring fragmentation. Planetesimal ablation rates are limited to approximately 1\% up to a core mass of 2 \me, whereas the pebbles are fully evaporated before 0.5 \me. Right: Panel (b) shows breakup distances for strong (100 MPa, red solid curve) and weak rock (1 MPa, red dashed curve). The black dashed curve follows the core radius. The crosses represent the points at which the planetesimals have enough time to radially spread to twice their initial radius. All plotted curves are almost identical for both enrichment assumptions (rainout plotted)
\label{breakup_ablation}}
\end{figure*}

In order to show the main differences in core growth between pebbles and planetesimals, we perform a further analysis on planetesimal mass deposition rates. We consider 1 km planetesimals for two reasons. Firstly, they serve as a limiting case of relatively severe thermal ablation compared to more massive impactors. Secondly, when planetesimals are sufficiently massive, they become dynamically stable due to their self-gravity. For rocky impactors this can happen when their radius exceeds about 70-100 km \citep[e.g.,][]{Podolak1988, Mordasini2015}. We use a typical planetesimal accretion rate of $10^{-6}$ \me $\textrm{yr}^{-1}$ \citep[e.g.,][]{Pollack1996}. Fig. \ref{breakup_ablation}a shows the mass fraction that 1 km planetesimals have lost upon core impact. Ablation rates remain limited to one percent even up to a core growth of 2 \me, which is consistent with previous calculations where planetesimals are assumed to reach the core intact.

Instead of ablation, the breakup of planetesimals during impact is typically expected to be the main cause of their mass deposition \citep{Podolak1988, Mordasini2015}. To calculate breakup points, we checked where in the envelope the dynamical pressure $P_\textrm{dyn} = \rho_\textrm{g} v_\textrm{i}^2$ first exceeds the compressive strength of the planetesimal $S_\textrm{c}$. Unfortunately, compressive strengths for planetesimals are largely unknown. Modeling of fragmentation events in the Earth's atmosphere and testing of the impactors post-impact yields compressive strengths in the broad range of 1-500 MPa \citep[e.g.,][]{Chyba1993, Svetsov1995, Petrovic2002, Popova2011, Podolak2015}. This large variation is due in part to differences in density, size, and the number of pre-existing faults in the rock. Generally, strength decreases with impactor size as the volume expands in which faults can occur \citep{Benz1999, Steward2011}. Around 100 m radius the opposite effect is true, and self-gravity effectively begins to strengthen the object. We use 1 km planetesimal strengths of 1 and 100 MPa for weak and strong rock, respectively. Compressive strength for ice is approximately an order of magnitude below this \citep{Petrovic2003, Podolak2015}, and gravitational strengthening, though it still occurs, is lessened by the lower density. The results for weak and strong rock are plotted in Fig. \ref{breakup_ablation}b, along with a curve to signify the core radius for comparison. In the weak limit, breakup can happen in envelopes with core masses as small as 0.2 \me. In contrast, strong 1 km planetesimals are found to be resilient to breakup until the core mass exceeds 1.2 \me.

Even if breakup occurs, it does not necessarily imply an instantaneous deposition of the planetesimal's mass in high-$Z$ vapor. The fragmentation event does not alter the kinetic energy of the post-breakup material until the shape of its bow shock has changed. There are various ways to model this, i.e., by assuming that the impactor flattens out as in the pancake model \citep{Mordasini2015} or that it expands radially, to subsequently split into smaller fragments when the radius of the post-breakup material becomes too large to support a common bow shock \citep{Hills1993}. We aim only to get an impression of the importance of this effect. Therefore, we choose a very simple model, similar to \citet{Hills1993}, where it is assumed that the dynamical pressure works to spread the post-impact material radially. Neglecting self-gravity and equating the work done by the dynamical pressure to the time derivative of the kinetic energy of the radially expanding cloud, we find that
\begin{equation}
P_\textrm{dyn} A_\textrm{i} \dot{R}_\textrm{i} = \frac{d}{dt} \left( \frac{3 M_\textrm{i} \dot{R}_\textrm{i}^2}{10} \right)
,\end{equation}
where $P_\textrm{dyn}$ is the dynamical pressure. This can be written as
\begin{align}\label{spreading}
\ddot{R}_\textrm{i} &= \left( \frac{5}{3} \right) \left( \frac{A_\textrm{i}}{M_\textrm{i}} \right) P_\textrm{dyn} ,\\
 &= \left( \frac{5}{4} \right) \left( \frac{P_\textrm{dyn}}{\rho_\textrm{i} R_\textrm{i}} \right), 
\end{align}
which is very similar to typical expressions of post-fragmentational spreading, the only difference being our factor $\frac{5}{4}$ compared to the use of a constant of order unity as in \citet{Mordasini2015}. To get an estimate of the velocity of radial spreading at the breakup point, we substitute $P_\textrm{dyn} = S_\textrm{c}$, to find
\begin{equation}
\ddot{R}_\textrm{i, breakup} = \left( \frac{5}{3} \right) \left( \frac{A_\textrm{i}}{M_\textrm{i}} \right) S_\textrm{c}
\end{equation}
The corresponding timescale to spread out the planetesimal is hence

\begin{equation}
    t_\textrm{spread} = \sqrt{\frac{R_\textrm{i}}{\ddot{R}_\textrm{i}}}
    \sim 10\, \mathrm{s} \left( \frac{R_\textrm{i}}{\mathrm{km}} \right) \left( \frac{S_\textrm{c}/\rho_\textrm{i}}{10^4\,\mathrm{m^2\,s^{-2}}} \right)^{-1/2}
    \label{eq:tspread}
,\end{equation}
which may be regarded as the time for a planetesimal's fragments to seperate. Once initiated, the spreading of the planetesimal will accelerate because of the linear dependence of $R_\textrm{i}$. We find that these timescales are small compared to the travel time of planetesimals between breakup and core impact. The crosses in Fig. \ref{breakup_ablation}b indicate the onset of this condition ($t_\textrm{break} +t_\textrm{spread} < t_\textrm{core}$). Therefore, (catastrophic) planetesimal fragmentation is likely, at least for small planetesimals. Sufficiently large planetesimals can be strengthened by self-gravity and may directly reach the core, in line with the standard assumption in the literature \citep[e.g.,][]{Podolak1988, Mordasini2015}.

Finally, even when planetesimals fragment and spread, it does not necessarily imply efficient deposition of high-$Z$ (silicate) vapor. When the fragmentation takes place at distances close to the core, which is usually the case (see Fig. \ref{breakup_ablation}b), the fragmentation is very localized and our spherically symmetric structure model would greatly overestimate the absorption rates. Indeed, as planetesimals are impacting at velocities close to the escape velocity, the impact cone will be very collimated. Then, it is likely that any high-Z vapor in the impact plume will quickly become supersaturated.
In conclusion, we envision that core growth by planetesimal accretion can proceed in a similar way to pebble accretion (with limited direct core growth, see Sect. \ref{Standard model}), provided that the impacting plantesimals are
\begin{enumerate}
\item
sufficiently weak (stronger plantesimals extend the duration of impacts, and thus direct core growth);
\item
sufficiently small (self-gravity will prevent efficient fragmentation beyond radii of 70-100 km);
\item
able to be absorbed efficiently by the envelope (localization of deposited mass in large fragmentation events can prevent this from happening).
\end{enumerate}

A more precise model, which is beyond the scope of this work, should account for these effects to investigate the standard assumptions of planetesimals reaching the core (see \citealt{Mordasini2015} for the present state of the art).
On the other hand, if we assume that planetesimals reach the core intact (barring ablation), we can simulate the planet's growth in exactly the same way as the pebble case. We then find that the envelope becomes dynamically unstable due to the accretion of disk gas when the core has grown to several \me, i.e., the critical core mass: 3.8 \me for the rainout case and 4.4 \me for mixing. \footnote{The critical core mass of the rainout scenario is lower here because the mean molecular weight is enhanced compared to the mixing case where only part of the ablated high-Z material is in vapor form. The rainout scenario amounts to a larger core mass in pebble accretion because this refers to direct core growth, not the critical core mass we refer to here.}

\subsection{Core surface conditions}\label{Core surface conditions}
\begin{figure}[h!] 
\centering
\includegraphics[width=\hsize]{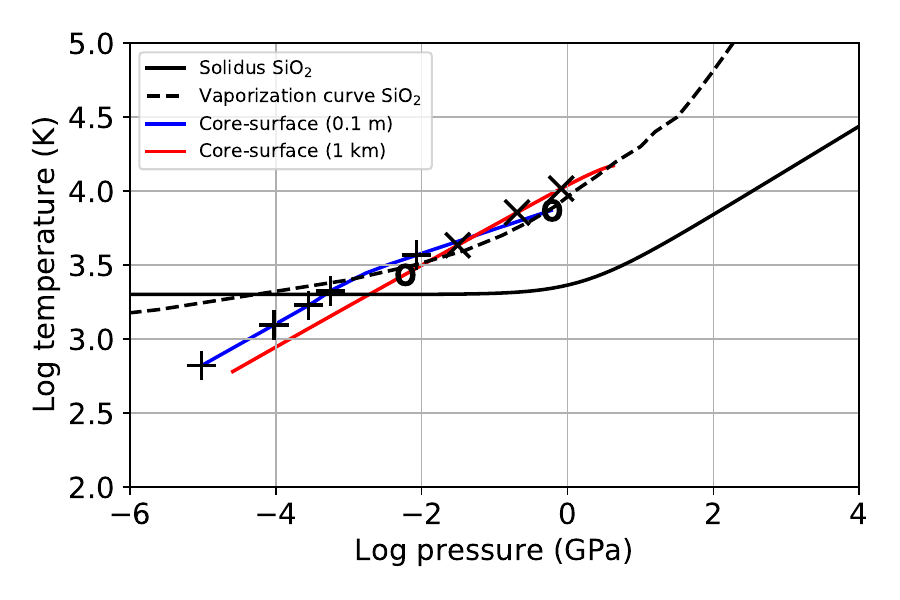} 
\caption{Core-surface conditions during core growth by pebble and (indestructible) planetesimal accretion. The core masses in the pebble (blue line) and planetesimal (red line) cases range from 0.1 \me to 0.59 \me and 3.8 \me, respectively. The pluses on the pebble line indicate 0.1 \me core-mass intervals; the crosses on the planetesimal line indicate 1 \me intervals. The location of the 0.59 \me point is indicated by the circles. The \sio2 solidus and QEOS vaporization curve are indicated by the solid and dashed lines, respectively.
\label{core_conditions}} 
\end{figure}

Another way to characterize the envelope of our core growth model is to study the core surface conditions. In Fig. \ref{core_conditions} we present a \sio2 phase-diagram with the quotidian equation of state (QEOS) as the vaporization curve, and the solidus from \citet{Poirier2000}. We use the more extensive QEOS here instead of the simple vapor pressure formulation from \citet{Stull1947} (Eq. 33) that we used in our simulations. This vapor pressure expression is valid for temperatures in the range up to 4000 K, and is therefore applicable to most of the envelope. On the other hand, the more extensive QEOS remains valid when P and T become large, conditions that are achieved at the core/envelope interface towards the end of our simulations. The figure shows comparison between 0.1 m pebbles accreting a core up to 0.59 \me, and indestructible planetesimals accreting up to 3.8 \me. Despite the difference in core mass between the pebble and planetesimal cases, we observe that the conditions at the core reach a pressure of 1 GPA and a temperature of $10^4$ K. In the pebble case, the high-Z inner region of the planet drives up the internal pressure and temperature, whereas in the planetesimal case the envelope contains a much more extended envelope with less enrichment: the envelope is never fully saturated with vapor, as only a small fraction of the impacting planetesimal mass ablates, and the rest reaches the core. Their initial difference is due to the higher accretion rate we use for pebbles, leading to increased luminosity (see Sect. \ref{planetesimal comparison}).
    
The core surface conditions are such that \sio2 passes through different phases during the simulated period. We provide a brief discussion for the pebble case. In the first phase ($M_\mathrm{core} \lesssim 0.35$ \me), pebbles reach the core unimpededly and the core remains completely solid. Soon thereafter, the vapor pressure becomes non-negligible at the core, and a fraction of the accreting \sio2 is absorbed as vapor. This is initially a small fraction, as the partial pressure of \sio2 is limited to the saturation curve set by the vapor pressure. After the core mass has grown beyond $ \gtrsim 0.45$ \me, the pressure at the core exceeds the vapor pressure (and the QEOS curve, as they are consistent in this regime), such that \sio2 becomes gaseous. In principle, the outer core will begin to partially evaporate under these P-T conditions. However, our model assumes that the envelope is in quasi-hydrostatic equilibrium and that the partial pressure of \sio2 is limited to the total pressure in the envelope, such that the fraction of \sio2 vapor will reach unity. We also assume that the enrichment is continually supplied by the pebble flux. Therefore, in our model the \sio2 at the core will only evaporate when the pebble flux is insufficient to fully saturate the envelope. In the simulated period, the envelope remains fully saturated by the pebble flux, meaning that the outer core cannot evaporate and continues to grow through high-Z rainout.

Our simulations end when all the accreting \sio2 pebble mass can be absorbed by the envelope and therefore direct core growth is halted. We find from the QEOS that the core-surface conditions at this point are on the edge between gaseous and liquid, such that the state of the core surface is uncertain. It is possible that the outer layer
of the core liquefies, or evaporates during the planet's subsequent evolution, similar to the scenario described by \citet{Chambers2017} for water-planets.

\section{Discussion and conclusions}\label{Discussion and conclusions}
Planet formation by pebble accretion is an alternative to the planetesimal-driven core accretion scenario. The two main differences are a higher expected solids accretion rate and more rapid ablation in the case of pebbles. In this work, we focused on the second difference and analyzed its effect on core growth. To accomplish this, we simulated the early formation of planets for both the pebble and planetesimal-driven core accretion scenarios, using a code consisting of two components: an impact model and a planet evolution model. We find that the rapid ablation of pebbles changes the process of core growth, which means that pebbles can be prevented from directly reaching the core of even a small growing protoplanet. Instead, their ablated material can either rain out to the core or be absorbed by the envelope as vapor. We have incorporated these enrichment effects from the moment that pebbles first start ablating by modifying the mean molecular weight and gas characteristics locally where high-Z material is present in vapor form.

We find that impacting \sio2 pebbles will fully ablate upon impact with planets of mass less than 0.5 \me. The amount of high-Z material that can stay in vapor form is heavily temperature dependent. This means that only the envelope's inner region is able to retain this material, leading to the formation of a high-Z layer surrounding the core where temperatures are highest. At first this region is small and unable to contain large amounts of vapor. Most ablated material then rains out to the core. Core growth slows down as the high-Z region expands outwards over time, increasing ablation for subsequent impactors and preventing their mass from reaching the core by absorbing it as vapor. This process limits direct core growth by pebbles to about 0.6 \me. Our findings are relatively insensitive to the planet's position in the accretion disk, pebble sizes up to 1m, or the solids accretion rate. They depend most sensitively on the mean molecular weight of the high-Z material and the adopted expression for the vapor pressure (see Eq. (\ref{vapor pressure})). The mean molecular weight determines the steepness of the temperature and density curves in the inner region, while vapor pressure determines its size and influences pebble ablation profiles.

By repeating our simulations with 1 km impacting planetesimals, we have shown that enrichment due to ablation is minimal if the impactors are sufficiently massive. Instead, breakup close to the core becomes an important mass deposition mechanism for impactors that are not sufficiently large to be held together by self-gravity. Plantesimals of 1 km are found to fragment when the planet's core mass exceeds 0.2 -- 1.2 \me, depending on their compressive strength. We estimate that breakup occurs sufficiently far away from the planetary core for the post-breakup material to separate, but the localization of this mass deposition likely still means that only a small portion of the mass can be absorbed by the envelope. This process should be studied in more detail in further work, as the ability of impactor mass to reach the core-envelope boundary is what determines the size to which planetary cores can grow. If efficient absorption of post-fragmentation mass is not prevented by its localization, only planetesimals large enough to prevent fragmentation can directly grow cores of several \me.

In our standard model, we use impactors with a uniform rocky (\sio2) composition. In reality, impactors outside the iceline constitute of a mix of rocky and icy pebbles. Pebbles made of water ice will evaporate farther away from the core than rocky ones, after which they can be absorbed by the atmosphere as vapor. We find that this happens before the core has grown to 0.1 \me. The inner envelope conditions are such that liquid water can form around the core, as described by \citet{Chambers2017}. In the planet's subsequent evolution, the \h2o vapor can either remain in the envelope where it increases the mean molecular weight and speeds up the envelope collapse \citep{Venturini2015} or it can recycle outwards into the disk \citep{Ormel2015, Cimerman2017, Lambrechts2017}. In both cases, \h2o does not contribute to any further core growth. Rocky materials require higher temperatures to vaporize and can therefore continue to grow the core. An implication of this finding is that pebble accretion is only able to form rocky cores beyond 0.1 \me, whereas icy cores must thus be formed by impacts of larger, planetesimal-sized impactors. It also offers an explanation for the assertion that close-in planets are predominantly rocky \citep{OwenWu2017,JinMordasini2017} based on an observed gap in their radius distribution \citep{Fulton2017}.

Recently, \citet{Alibert2017} has suggested that, if recycling operates vigorously and the envelope becomes fully mixed, any high-Z material present in the envelope will recycle back to the disk. In that case, direct core growth is the only growth mechanism in pebble accretion, core masses would be limited to 0.6 \me, and enrichment of the envelope would be very low. However, there are two arguments against his hypothesis of efficient mixing. First, recent hydrodynamical simulations involving radiation transport have shown that hydrodynamical recycling operates less efficiently in the inner-most regions where we expect the high-Z vapors to be concentrated \citep{Cimerman2017, Lambrechts2017}. Second, the high-Z vapor will congeal when it is transported to cooler regions. These grains will subsequently rain out to the core, unless they can be very efficiently mixed with the outer envelope, i.e., by regular (eddy) convection. Conceivably, luminosity sources from radioactive heating or core cooling (Vazan et al. 2017 in prep.) could provide the required luminosity.

However, this mixing is only effective if the planet contains a large convective region. In our work, we find that the convective region is comparatively small and shielded by a large isothermal, radiative zone (see Fig. 4). Furthermore, in order to mix grains into the outer envelope, the (convective) mixing times must be shorter than the settling timescale ($\tau_\textrm{mix} <\tau_\textrm{settle}$). Given that the high-Z layer is situated directly on top of the core, settling timescales are very short even for submicron-sized grains. Therefore, we argue that the planet is likely to retain its high-Z vapor against recycling. 

Under these conditions the formation of a high-Z layer, situated between the core and the H/He envelope, is an integral part of planet formation by pebble accretion. It can be considered as a dilute extension of the core, with densities approaching the core density and a metal-poor surrounding envelope. Upon further accretion of pebbles, the high-Z region will keep expanding outwards, increasing in mass along with the H/He-region. As it does in the planetesimal-driven scenario, this will end when hydrostatic balance can no longer be maintained. However, accretion of pebbles may already terminate before this point. In that case the protoplanet will evolve through Kelvin--Helmholtz cooling of the envelope \citep[e.g.,][]{LeeEtal2014,ColemanEtal2017}. Because of the cooling, we then expect the high-Z layer to rain out to the core. This is the way we envision core growth to super-Earth and mini-Neptune sizes by pebble accretion.

To summarize, we find that pebble accretion can only directly form rocky cores up to 0.6 \me, and is unable to form icy cores larger than 0.1 \me. This contrasts with planetesimal-driven core accretion, which can directly produce more massive cores with various compositions if the planetesimals are sufficiently strong and large to deposit their mass at the core as is, hitherto, the standard approach in the literature. The reason for limited direct core growth in pebble accretion is that pebbles ablate more rapidly upon impact. Their subsequent absorption by the surrounding gas prevents their mass from reaching the core when the envelope becomes sufficiently hot and massive. Core growth after this point may proceed through new indirect processes if the planet is able to retain its high-Z material. The expected localization of \sio2 vapors near the core can prevent it from being recycled into the disk and facilitate future core growth when the planet cools down. Further research is required to explore these later stages of core growth, and we will study it in a future work.

\section*{Acknowledgements}
\tiny{C.W.O. is supported by The Netherlands Organization for Scientific Research (NWO; VIDI project 639.042.422). We thank the referee, Michiel Lambrechts, for a very constructive report.}

\bibliographystyle{aa} 
\bibliography{core_growth_bib_AA}

\end{document}